\documentclass[useAMS,usenatbib]{mn2e}
\usepackage{graphicx}

\usepackage{subfigure}
\usepackage{amssymb}
\usepackage{epstopdf}
\usepackage{rotating}
\usepackage{amsmath}
\usepackage{color}

\title[Growth of BCGs via mergers since $z=1$]{Growth of brightest cluster galaxies via mergers since $z=1$}
\author[C. Burke \& C. A. Collins]{Claire Burke$^{1}$\thanks{E-mail: cb@astro.livjm.ac.uk (C.B.)} \& Chris A. Collins$^{1}$\\
$^{1}$Astrophysics Research Institute, Liverpool John Moores University, IC2, Liverpool Science Park, 146 Brownlow Hill, \\~~~Liverpool, L3 5RF, UK.}
\begin{document}

\date{Accepted. Received ; in original form }

\pagerange{\pageref{firstpage}--\pageref{lastpage}} \pubyear{2011}

\maketitle

\label{firstpage}

\begin{abstract}

Hierarchical assembly within clusters of galaxies is tied directly to the evolution of the Brightest Cluster Galaxies (BCGs), which dominate the stellar light in the centres of rich clusters. In this paper we investigate the number of mergers onto BCGs in 14 X-ray selected clusters over the redshift range $0.8<z<1.4$ using HST imaging data. We find significant differences in the numbers of companion galaxies to BCGs between the clusters in our sample indicating that BCGs in similar mass clusters can have very different merging histories. Within a 50~kpc radius around the BCGs we find an average of $6.45\pm1.15$ companion galaxies with mass ratios (companion:BCG) between 1:1 and 1:20. The infalling companions show a 50/50 split between major (1:1 - 1:2) and minor (1:3--1:20) mergers. When compared to similar work using lower redshift clusters, these results demonstrate that both major and minor merging was more common in the past. Since the dynamical timescales for merging onto the BCG are relatively short compared with the look-back time to $z\sim1$ our results suggest that the BCG stellar mass may increase by as much as 1.8 times since $z=1$.  However the growth rate of BCGs will be substantially less if stripped material from nearby companions ends up in the diffuse intracluster light.

\end{abstract}


\section{Introduction}\label{intro}

Brightest cluster galaxies (BCGs) are the most massive and most homogeneous class of galaxy observed in the Universe. Their large stellar luminosities and unique position at the centres of galaxy clusters make them easily identified from cluster surveys and observed out to high redshift. Occupying the dense regions in cluster cores, the formation of BCGs can be explained by dynamical friction induced major merging (White 1976; Ostriker \& Hausman 1977). The same general picture subsequently emerges from simulations which show that BCGs could form through dissipationless merging of similar-mass systems (e.g. Dubinski, Mihos \& Hernquist, 1999). More generally, work on elliptical galaxies from semi-analytic simulations indicates that they originate from an early period of rapid star formation and major merging at $z>2$ with subsequent mass assembly becoming dominated by non-dissipative (dry) minor merging (e.g. Khochfar \& Silk 2006) -- a conclusion which fits reasonably well with the ``downsizing'' behaviour results from deep optical/IR surveys (e.g. Cirasuolo et al. 2010). Since BCGs are formed in the first density peaks to collapse their evolution is easily followed in large $\Lambda$CDM cosmological simulations of structure formation. Based on the semi-analytic modelling of the dark matter Millennium Simulation, for example, De Lucia \& Blaizot (2007) predict that BCGs have only $20\%-30\%$ of their stellar mass assembled by $z\sim1$ and should be undergoing many mergers between $z=1-0$, approximately quadrupling in stellar mass over this time period. The recent simulations of Laporte et al. (2013) predict a somewhat smaller stellar mass growth for BCGs by both major and minor mergers, with mass growth of a factor of 1.9 during the time between $0.3 < z < 1.0$ and a further factor of 1.5 since $z=0.3$. Observational studies of the merging rates in the centres of clusters concentrate on samples at low redshift, concluding 
that major mergers of BCGs are a rare occurrence and reporting a dominance of small companion galaxies close to the BCGs in cluster cores (Edwards \& Patton, 2012; Liu et al., 2009; McIntosh et al. 2008). However, other authors point out that BCGs at low redshift continue to grow from major merging (e.g. Brough et al. 2011).

The close relationship between the complex intracluster environment and the central BCG is well established (e.g. Edge 1991; Collins \& Mann 1998; Brough et al. 2005; Stott 2008, 2012). But despite the obvious consequence of BCG growth in mass via the merging processes in the centres of clusters, observationally the problem remains a vexing issue. The claim by Arag\'{o}n-Salamanca, Baugh \& Kauffmann (1998) of evolution in the stellar mass using the observed near-IR brightness of BCGs was shown to result from optical selection bias (Burke, Collins \& Mann, 2000). Subsequent work indicated little evolution of BCG stellar mass in both optically selected clusters out to $z=0.8$ (Whiley et al. 2008) and also for BCGs based on X-ray cluster selection reaching to $z=1.5$ (Collins et al. 2009, Stott et al. 2010). The evolution measured for the BCGs in X-ray clusters as far back as $z\sim1$ is surprisingly small, with these systems already having of $90-95\%$ of their final mass and $\sim70\%$ of their final scale-size (Stott et al. 2011) compared to their present day counterparts.  On the other hand Lidman et al. (2012) claim to measure a stellar mass growth of a factor of 1.8 over the same epoch based on 12 BCGs from the Spitzer-based SpARCS survey. This result is significantly different from the X-ray based clusters despite the substantial overlap in redshift of the samples, and while the SpARCS BCG data are still a factor 1.5 heavier than the predicted masses of De Lucia \& Blaizot (2007) they are in better agreement with the BCG growth rates from the most recent simulations (Laporte et al. 2013).

Independently of these considerations, it is becoming increasingly clear that to make any meaningful comparison between observed and predicted BCG evolution it is essential to take into account the stellar mass growth of the diffuse intracluster light (ICL) which extends well beyond the stellar profile of the BCG in most cases but remains almost degenerate with the underlying surface brightness profile of the BCG in the inner $50-100$\,kpc. At low redshift both observations (e.g. Gonzalez, Zaritsky \& Zabludoff, 2005, 2007; Zibetti et al., 2005; Toledo et al., 2011) and simulations (Conroy et al., 2007) find that the light from the ICL and BCG combined contributes somewhere between $30-90\%$ of the total cluster starlight, with the individual ICL contribution similar to or even dominating (up to $80\%$) the contribution from the BCG. 
Furthermore the results of Burke et al. (2012) show that whatever may be happening to the BCG as a result of mergers, the ICL component is built up relatively recently, increasing by as much as a factor 4 between $z=1$ and the present. We return to this issue later in the paper.  

In this paper we address the issue of BCG assembly using $HST$ `snapshots' to estimate the instantaneous number of major and minor mergers onto BCGs from a sample of 14 high-redshift X-ray luminous clusters, and then compare our results to those of low redshift studies and simulations.
      
The structure of this paper is as follows: Section~\ref{data_s} contains the details of the sample studied; in Section~\ref{method_s} the methods and corrections used are described; in Section~\ref{results_s} the numbers of BCG companions and the mass increases from their accretion are presented; the results are discussed in Section~\ref{discussion} and conclusions drawn in Section~\ref{conc_s}.
Throughout this paper we adopt a $\Lambda$CDM cosmology with $H_0$ = 70 km s$^{-1}$Mpc$^{-1}$, $\Omega_M$ = 0.3, $\Omega_{\Lambda}$ = 0.7.

\section{A high redshift sample of BCGs}\label{data_s}

We construct a sample of 14 X-ray selected galaxy clusters at $0.8<z<1.4$, all of which have $I$-band imaging data from {\it HST Advanced Camera for Surveys (ACS)} or {\it Wide Field Planetary Camera 2 (WFPC2)}. The high spatial resolution of the $HST$ enables BCG companions and galaxy separations to be resolved to a few kpc at $z=1$.

\begin{figure*}
\centering
\includegraphics[]{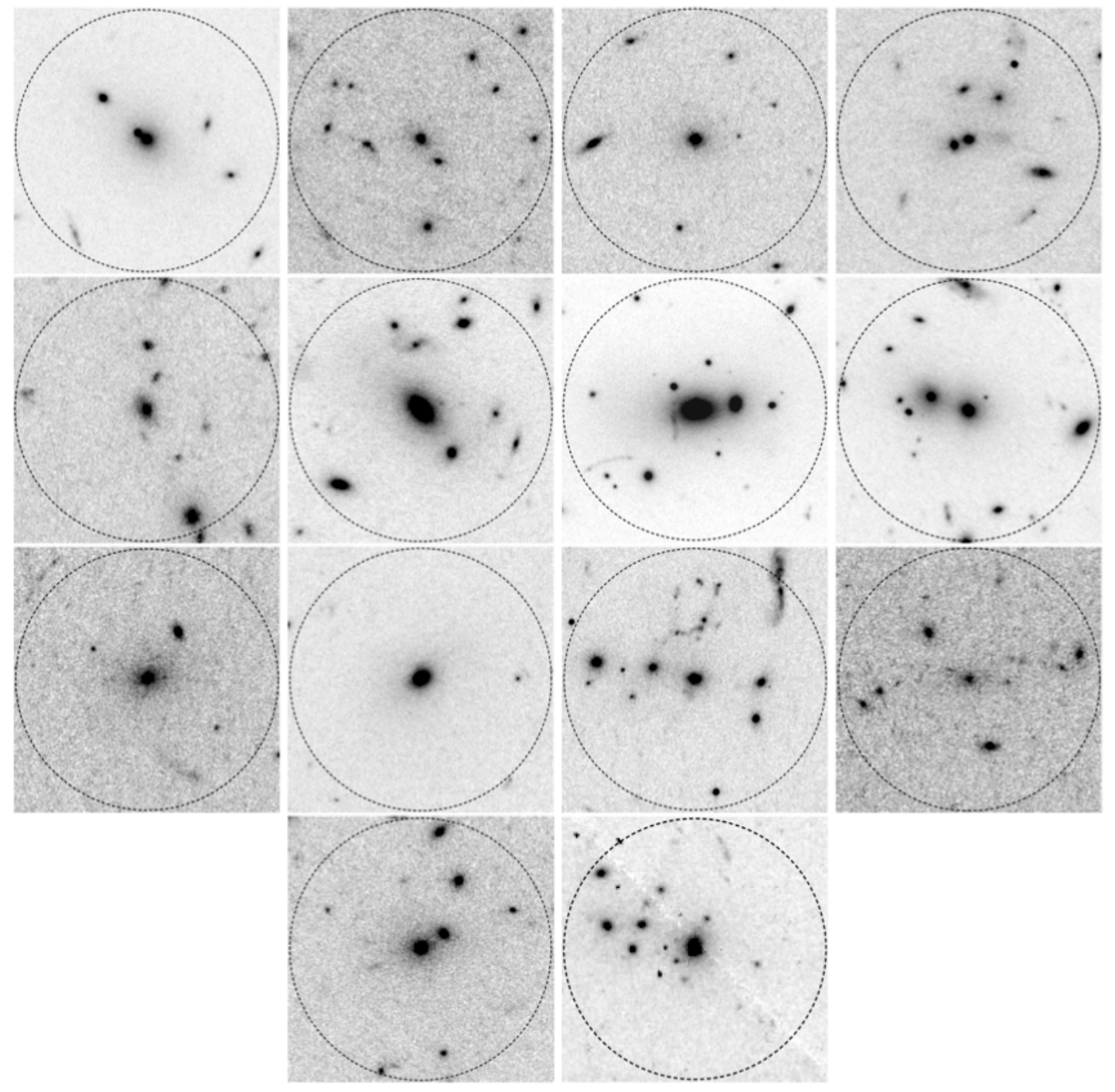}
\caption{The inner 50 kpc radius region surrounding the BCGs of the clusters in this sample. Images are 100 kpc $\times$ 100 kpc at the redshift of the cluster shown. Clusters are ordered from top to bottom and left to right in the same order as in Table~\ref{clusters}. Dotted lines indicate circle of projected 50 kpc radius.}
\label{bcgs_50kpc}
\end{figure*}

The majority of the sample is drawn from that of Stott et al. (2010) which are all X-ray selected clusters, and was supplemented with 2 additional X-ray luminous clusters in the same redshift range (RCS J2319.9+0038, Gilbank et al., 2008; and RX J1716.4+6708, Gioia et al., 1999) with similar $HST$ $I$-band data. The clusters in the sample have images taken with either the F850LP or F814W filter, both of which correspond to observed frame $I$-band or rest frame $\sim B$-band at the redshift of our sample. The data are downloaded from the $HST$ archive in pre-reduced form. The details of the clusters in this sample are shown in Table~\ref{clusters} and Figure~\ref{bcgs_50kpc} shows the central 100\,kpc region surrounding all 14 clusters.

\begin{table*}
\centering
\caption{The galaxy clusters in this sample and relevant observation details. All clusters were observed in the $I$-band with $HST$. Cluster $M_{200}$ masses are taken from Stott et al. (2010) with the exception of RCS J2319.9, which is taken from Gilbank et al. (2008), and RX J1716.4 which is taken from Gioia et al. (1999).}
\begin{tabular}{l c c c c c c c}
\hline
Cluster name		& RA	 		& Dec 	& Redshift 	& Observing 	& Exposure	& Zero 	& Cluster mass \\
			&			&		&			&  instrument/filter	& time (s)		& point	& $\times 10^{1	4}M_{\odot}$\\
\hline
Cl 0152		& 01 52 41.0	& -13 57 45	& 0.83	& ACS/F850LP		& 19000	& 24.351		& ~4.5 \\
XLSS 0223	& 02 23 53.9	& -04 36 22	& 1.22 	& ACS/F850LP		& 2000	& 24.351		& ~1.8 \\
RCS 0439	& 04 39 38.0	& -29 04 55	& 0.95	& ACS/F850LP		& 2000	& 24.351		& ~0.5 \\
RX0848		& 08 48 56.3	& ~44 52 16 	& 1.26	& ACS/F850LP		& 20900	& 24.350		& ~4.7 \\
RDCS 0910	& 09 10 44.9	& ~54 22 09	& 1.11	& ACS/F850LP		& 11440	& 24.351		& ~5.3 \\
MS1054		& 10 57 00.2	& -03 37 23	& 0.82	& ACS/F850LP		& 17760	& 24.351		& ~8.5 \\
Cl1226		& 12 26 58.0	& ~33 32 54	& 0.89	& ACS/F814W		& 24000	& 25.532		& 13.9 \\
RDCS 1252	& 12 52 54.4	& -29 27 17	& 1.24	& ACS/F850LP		& 76800	& 24.351		& ~6.1 \\
WARPS 1415	& 14 15 11.1	& ~36 12 03	& 1.03	& ACS/F850LP		& 1500	& 24.350		& ~5.2 \\
CL1604		& 16 04 25.2	& ~43 04 53	& 0.90	& ACS/F814W		& 4840	& 25.532		& ~1.2 \\
RCS 162009	& 16 20 09.4	& ~29 29 26	& 0.87	& ACS/F814W		& 1440	& 24.350		& ~3.4 \\
XMMU 2235	& 22 35 20.6	& -25 57 42	& 1.39	& ACS/F850LP		& 2000	& 24.351		& ~7.7 \\
RCS J2319.9+0038	& 23 19 53.3	& ~00 38 13 & 0.90 	& ACS/F850LP		& 1360	& 23.326		& ~6.4\\
RX J1716.4+6708	& 17 16 49.6	& ~67 08 30 & 0.81 	& WFPC2/F814W		& 2700	& 21.665	& ~2.0 \\

\hline
\end{tabular}
\label{clusters}
\end{table*}

\section{Measuring merger rates}\label{method_s} 

In order to measure the numbers of mergers of cluster galaxies onto the BCGs between $z=1$ and the present we follow a method similar to that adopted by both Edwards \& Patton (2012) for clusters at $z\sim0.3$ and the simulations of Laporte et al. (2013), with which we wish to compare our results. Therefore, to be consistent, we also measure the number of companions to each BCG inside an aperture of 50 kpc radius. This choice is otherwise somewhat arbitrary, but justified a posteriori as the dynamical timescale for merging at this radius is less than the cosmic time between $z=1$ and the present (an important criteria for our analysis, see Section~\ref{timescale}), while at the same time it is large enough to encompass the typical half-light radii of our BCGs (Stott et al. 2011) and thus contain the majority of the BCG stellar light (Burke et al. 2012).

The apparent magnitude of the BCG and its companions within the 50\,kpc aperture are measured with {\tt SExtractor} (Bertin \& Arnouts, 1996), using the {\tt MAG\_AUTO} method in the same manner as described in Stott et al. (2010) and Lidman et al. (2012). Any point sources found inside the 50\,kpc apertures are excluded from the counted companions; point sources are classified as objets with a stellar light profile similar to that of the PSF in the data. 

The luminosity ratio of BCG to companion $L_{BCG}:L_{comp}$ is determined from their apparent magnitude ($m$) difference by,
\begin{equation}
m_{BCG} - m_{comp}= - 2.5log_{10}\left(\frac{L_{BCG}}{L_{comp}}\right).
\label{lum_equ}
\end{equation}

In line with Edwards \& Patton (2012) and  Laporte et al. (2013), any companion with a luminosity ratio to the BCG in the range 1:1 -- 1:2 is defined as a major merger, while those with luminosity ratios greater than 1:2 are defined as minor mergers.

\subsection{Completeness limit of companions}
\begin{figure*}
\centering
\includegraphics[width=6cm]{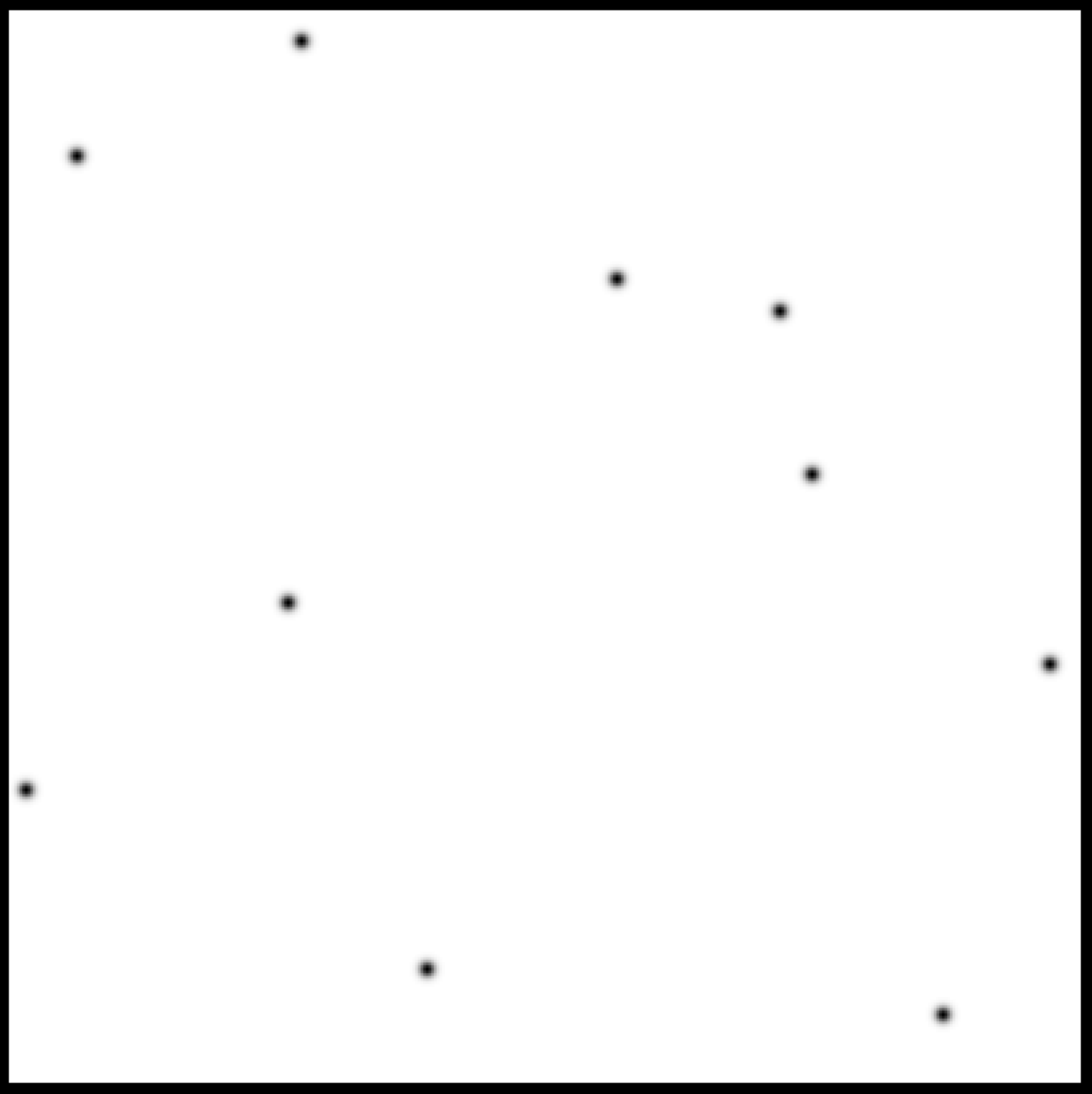}
\includegraphics[width=6cm]{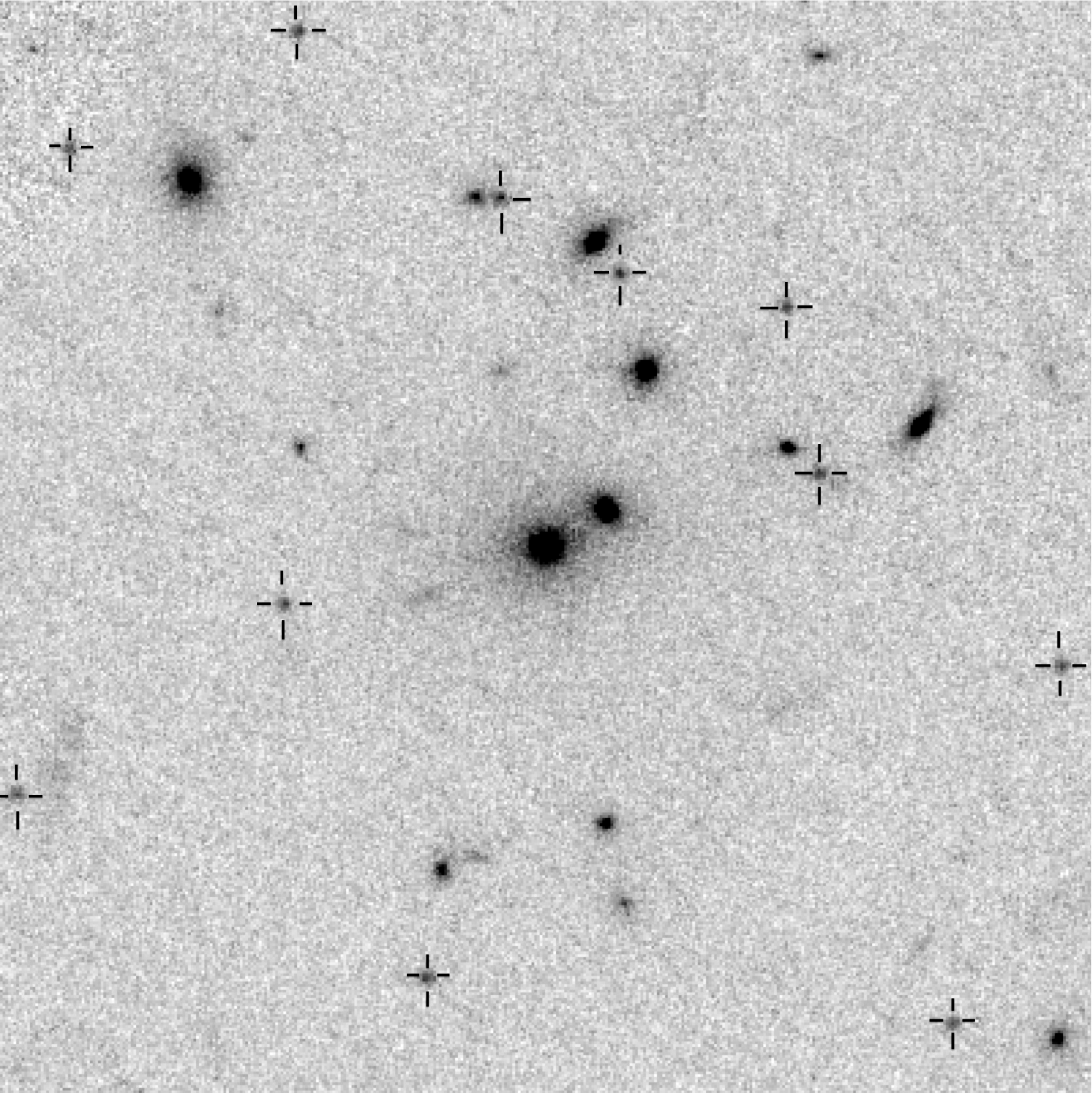}
\caption{Left: An example of the model PSFs which were produced to determine the magnitude completeness limit of the data. Right: the model PSFs placed into a cluster image, their locations are indicated by the cross-hairs. It can be seen that the model PSFs are of similar angular size to the smaller companion galaxies to the BCG. Images are $\sim$ 150 $\times$ 150 kpc.} 
\label{psf_mods}
\end{figure*}

To ensure that the companion galaxies are counted to a consistent limiting magnitude for all BCGs we estimate the completeness magnitude,  defined here as the faintest magnitude at which a source will be consistently detected. To do this a suite of simulated objects is combined with the observational data. Model $HST$ PSFs are made using {\tt TinyTim} (Krist, Hook \& Stoehr, 2011) with a FWHM of 0.5 arcsec, which is a similar angular size to the smaller companion galaxies of the BCGs in the sample. The model PSFs are generated with integer value apparent magnitudes and placed in the real cluster data at ten randomly generated spacial positions (excluding the inner 50$\times 50$\,kpc around the BCG). Ten random positions are chosen to account for possible variation in the background level of the image and the possibility of a model PSF being placed on top of another real object in the images. From the ten models we determine the faintest magnitude limit at which companion detection is at least 90\% complete. An example of the model PSFs and their positions is shown in Figure~\ref{psf_mods}.

After running {\tt SExtractor} in the same manner as described above to detect all sources in each field with multiple PSF profiles inserted, the recovered number of PSF sources is recorded. This process is repeated, combining PSFs of fainter apparent magnitude each time (decreasing in half-integer values) until the number of model PSFs recovered in the {\tt SExtractor} output falls below ten. The lowest magnitude for which all the model PSFs are recovered is taken as the companion galaxy completeness limit and these are shown for all clusters in Table~\ref{comp_tab}. Most of the data are complete to a luminosity ratio significantly greater than 1:20 (magnitude difference $<-3.25$) and for   the measurements made in this study we limit the companion galaxies to a luminosity ratio of $L_{comp}:L_{BCG}=$ 1:20 and exclude any companion objects measured below this luminosity limit.  The two clusters with the shallowest limits are XLSS~0223 and XMMU~2235, with  magnitude gaps $m_{BCG}-m_{comp} =-3.07$ (luminosity ratio 1:17) and $-3.14$ (luminosity ratio 1:18) respectively, and these just fall short of the cut but are still $90\%$ complete at 1:20 and so are retained in the sample. Objects with luminosity ratios (and hence mass ratios) greater than 1:20 of the BCG would have to be in large numbers to make a significant contribution to the mass growth in cluster cores, and in Section~\ref{timescale} we also demonstrate that there is insufficient time for companions of very large mass ratio with the BCG ($\mathcal{M}_{comp}:\mathcal{M}_{BCG}\sim$1:100) at $z=1$ to be accreted onto the BCG by the present day, so these small mass objects will not contribute significantly to the mass growth of the BCG.

\begin{table*}
\centering
\caption{Magnitude completeness of the data for each cluster, indicated by the maximum magnitude gap between the BCG and the faintest companion which can be consistently detected, and the equivalent luminosity ratio ($L_{comp}:L_{BCG}$) of the completeness limit.}
\begin{tabular}{l c c c}
\hline
Cluster name 	& $m_{BCG}$		&Completeness   		& Luminosity ratio\\
		      	& 		 	&$m_{BCG}-m_{comp}$		&$L_{comp}:L_{BCG}$\\
\hline
Cl 0152		&20.44 $\pm$ 0.18			&-5.06			& 1:106\\
XLSS 0223	&21.33 $\pm$ 0.27			&-3.07			& 1:17\\
RCS 0439	&20.45 $\pm$ 0.18			&-4.05			& 1:42\\
RX0848		&21.96 $\pm$ 0.36			&-3.44			&1:24\\
RDCS 0910	&21.94 $\pm$ 0.36			&-3.56			&1:27\\
MS1054		&19.14 $\pm$ 0.10			&-5.86			& 1:221\\
Cl1226		&19.55 $\pm$ 0.07			&-6.97			& 1:615\\
RDCS 1252	&20.98 $\pm$ 0.23			&-5.53			& 1:163\\
WARPS 1415	&19.98 $\pm$ 0.15			&-4.52			&1:64\\
CL1604		&19.63 $\pm$ 0.11			&-6.38			& 1:355\\
RCS 162009	&19.77 $\pm$ 0.13			&-4.23			&1:49\\
XMMU 2235	&22.06 $\pm$ 0.38			&-3.14			&1:18\\
RCS J2319.9+0038	&18.59 $\pm$ 0.12		&-4.91			&1:92\\
RX J1716.4+6708	&17.42 $\pm$ 0.15		&-5.08			& 1:108\\

\hline
\end{tabular}
\label{comp_tab}
\end{table*}

\subsection{Comparison sample and contamination}
In order to account for the contamination from projected foreground or background galaxies in the measured BCG companion counts a comparison sample of field galaxies is created by searching each cluster image for elliptical galaxies with magnitudes within $\pm0.5$\,mag of the BCG in that cluster and positioned at the outskirts (radii from BCG $\gtrsim 1$ Mpc) of each cluster or in the field. This generates a comparison sample of 39 field ellipticals for which we carry out companion counts inside 50~kpc in an identical manner to that of the BCGs.

The overall contamination level from non-cluster galaxies is estimated at around $18\%$, a level which is consistent with that found (12\%) from the halo occupation distribution analysis of Capozzi et al. (2012) which was carried out on a sample containing 7 of the clusters used here. The largest contamination fraction is $22\%$ for the luminosity ratio bin $L_{comp}:L_{BCG} =$ 1:2 -- 1:5. The lowest contamination fraction is $7\%$, found for the lowest luminosity ratio bin ($L_{comp}:L_{BCG} =$ 1:1 -- 1:2). As expected this indicates fewer close pairs of similar brightness elliptical galaxies in the field than in the centres of clusters.

\section{Results}\label{results_s}

\begin{figure}
\centering
\includegraphics[width=8cm]{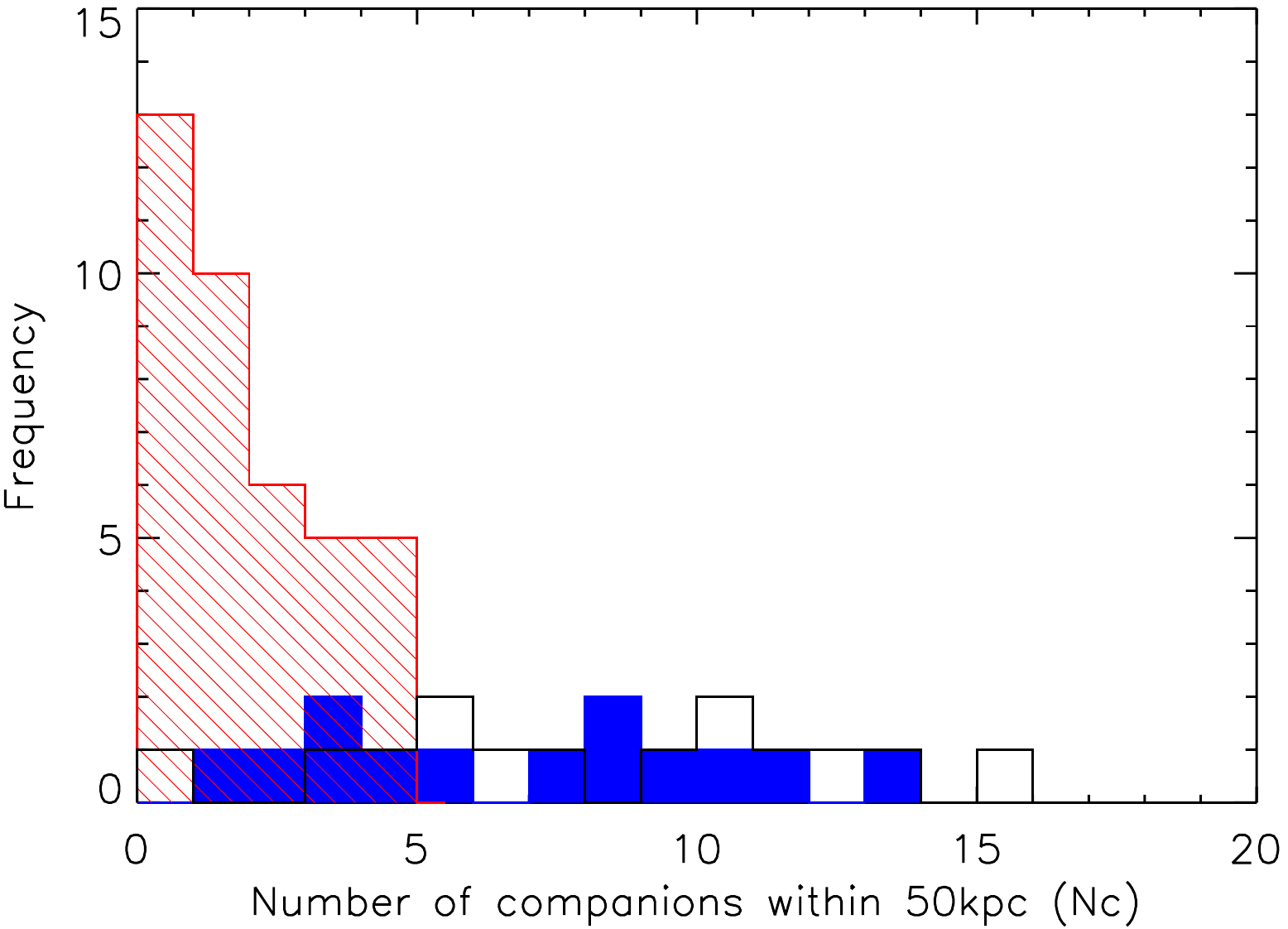}
\caption{Uncorrected numbers of companions for BCGs (black empty  histogram); numbers of companions for field comparison sample (red diagonal striped histogram); numbers of companions for BCGs after correction for contamination (blue filled histogram).}
\label{n_companions}
\end{figure}

The data for the companion counts ($N_c$) around each of the 14 BCGs; also broken down into luminosity ratio bins, along with the field averages and corrected companion counts are all shown in Table~\ref{results_tab} and Figure~\ref{n_companions}. 

The average number of companion galaxies to the BCGs within 50~kpc is $7.86\pm1.14$ and the estimated number of contamination companions is $1.41\pm0.19$ giving a corrected value for $N_c$ of 6.45$\pm$1.15.
 
\begin{table*}
\centering
\caption{Numbers of BCG companions, not corrected for contamination, and the mass increase inferred if all the companions merge onto the BCG by $z=0$. The average mass increases are found from the average of the minimum and maximum mass growth inferred from the luminosity ratio bins shown, described in Section~\ref{timescale} (see Equations~\ref{equ_min} and \ref{equ_max}).}
\begin{tabular}{l c c c c c c c c}
\hline
Cluster			&$z$		&$N_c$&	\multicolumn{4}{c}{$N_c$}			& \multicolumn{2}{c}{$\mathcal{M}_f$}\\
				&&&\multicolumn{4}{c}{Luminosity ratio ($L_{companion}:L_{BCG}$)}	&Mass increase& Mass increase\\
				&		&&1:1--1:2	&1:2--1:5	&1:5--1:10	&1:10--1:20	& (average)	& (corrected)\\
\hline

Cl 0152			&0.83	&5	&1	&2	&1	&1	&1.68 $\pm$ 0.98	&1.36	$\pm$	0.73\\
XLSS 0223		& 1.22	&10	&1	&4	&4	&1	&2.83 $\pm$ 1.55	&2.51	$\pm$	1.30\\
RCS 0439		&0.95	&4	&0	&1	&0	&3	&0.58 $\pm$ 0.36	&0.26	$\pm$	0.41\\
RX0848			&1.26	&11	&3	&2	&2	&4	&3.55 $\pm$ 1.95	&3.23	$\pm$	1.70\\
RDCS 0910		&1.11	&5	&0	&1	&0	&4	&0.65 $\pm$ 0.52	&0.33	$\pm$	0.47\\
MS1054			&0.82	&7	&0	&1	&3	&3	&1.03 $\pm$ 0.68	&0.71	$\pm$	0.43\\
Cl1226			&0.89	&6	&0	&1	&1	&4	&0.80 $\pm$ 0.66	&0.48	$\pm$	0.42\\
RDCS 1252		&1.24	&13	&3	&0	&6	&4	&3.45 $\pm$ 1.89	&3.13	$\pm$	1.75\\
WARPS 1415		&1.03	&3	&0	&1	&0	&2	&0.50 $\pm$ 0.42	&0.18	$\pm$	0.22\\
CL1604			&0.90	&0	&0	&0	&0	&0	&0.00 $\pm$ 0.00	&-0.32	$\pm$	0.12\\
RCS 162009		&0.87	&12	&1	&4	&2	&5	&2.83 $\pm$ 1.73	&2.51	$\pm$	1.48\\
XMMU 2235		&1.39	&15	&3	&7	&4	&1	&5.38 $\pm$ 2.89	&5.06	$\pm$	2.65\\
RCS J2319.9+0038	&0.90	&10	&3	&0	&3	&4	&3.00 $\pm$ 1.59	&2.68	$\pm$	1.34\\
RX J1716.4+6708 	&0.81	&9	&4	&1	&2	&2	&3.80 $\pm$ 1.90	&3.48	$\pm$	1.65\\

\hline
Uncorrected average	&	&7.86 $\pm$ 1.14	&1.36 $\pm$ 0.40	&1.79 $\pm$ 0.53	&2.00 $\pm$ 0.49	&2.71 $\pm$ 0.41 &2.15 $\pm$ 0.43 & \\
Field average	&			&1.41 $\pm$ 0.19	&0.10 $\pm$ 0.06	&0.41 $\pm$ 0.10	&0.41 $\pm$ 0.10	&0.48 $\pm$ 0.12		&\\
\hline
Corrected average 		&	&6.45 $\pm$ 1.15	&1.26 $\pm$ 0.40	&1.38 $\pm$ 0.54	&1.59 $\pm$ 0.50	&2.23 $\pm$ 0.43	& &1.83 $\pm$ 0.43\\
\hline

\end{tabular}
\label{results_tab}
\end{table*}

\subsection{Mass growth of BCGs since $z$=1}\label{timescale}
To calculate an estimate for the mass growth of the BCGs in the sample from the accretion of their companions, we must first ascertain whether the companions will merge with the BCG during the interval between $z=1$ and $z=0$. To do this we estimate the dynamical friction timescale defined as the time taken for a companion galaxy at a given radius to spiral into the centre of a larger galaxy's gravitational potential. Assuming the companions are in circular orbits around the BCG and the dark matter density profile of the BCG has a $r^{-2}$ shape (a singular isothermal sphere), the dynamical friction timescale, $T_{fric}$, is given by 
\begin{equation}
T_{fric}= 1.17\frac{r^2v_c}{G\mathcal{M}_{c}ln(\Lambda)}
\end{equation}
(Binney \& Tremaine, 1987), where $r$ is the initial separation of the galaxies, $v_c$ is the circular velocity,  $\mathcal{M}_c$ is the mass of the companion and $ln(\Lambda)$ is the Coulomb logarithm.
For $r$ in kpc, $v_c$ in km\,s$^{-1}$, $\mathcal{M}_c$ in solar masses and $ln(\Lambda)$= 2 (as measured for example in the parameter study of halo mass profiles by Dubinski et al. (1999) for interacting galaxies), $T_{fric}$ can be estimated in Gyrs using the expression 

\begin{equation}
T_{fric}= 1.32\times10^5 ~ \frac{r^2v_c}{\mathcal{M}_{c}}.
\label{tf_equ}
\end{equation}

The stellar masses of most of the BCGs in the sample are taken from Stott et al. (2010) who measure $K$-band magnitudes with {\tt SExtractor} using the {\tt MAG\_AUTO} method, which are close to the measured photometric values in a 50~kpc aperture. From these values the stellar masses of the companions can be inferred, assuming that their stellar luminosity traces their stellar mass in the same ratio as the BCG. The core regions of galaxy clusters are found to be overwhelmingly dominated by early type, red-sequence galaxies (e.g., Gladders \& Yee, 2000 and references within) so only elliptical galaxies are considered for this calculation. The total mass of the companion, $\mathcal{M}_{comp}$, is estimated from mass-to-light ratios of elliptical galaxies found in the literature, some examples of which are listed here. Van der Marel (1991) find an average mass-to-light for elliptical galaxies in the $B$-band of M/L$=5.93\pm$0.25 for absolute magnitudes $M_B\sim -20$ to $-23$, while Gerhard et al. (2001) find mass-to-light ratios for elliptical galaxies as large as 10 in the $B$-band with average absolute magnitudes $M_B\sim-21$. The observed frame $I$-band of the clusters examined here corresponds to rest frame $B$-band, and the range of absolute magnitudes from the previous studies described above is similar to that of the companion galaxies in this study. Taking a conservative assumption about the evolution of M/L, we adopt a value of M/L$=4$. This value is consistent with the typical M/L values ($4-6$) found in the modelling of the stellar evolution of early type galaxies at $z\sim1$ and the ICL (see Burke et al., 2012).

It is a ubiquitous, but often unjustified assumption, that the companions to BCGs are in circular orbits. In fact N-body simulations of merging objects (e.g. van den Bosch et al., 1999) show that objects on eccentric orbits have shorter dynamical friction timescales than those on circular orbits, with the timescale decreasing with increasing eccentricity of the orbits. In this sense our calculated dynamical friction timescale estimates will be upper limits, which are sufficient for our purposes. To estimate the circular velocity ($v_c$) of the companions, we use the relation between $v_c$ and velocity dispersion ($\sigma$) of the companions given by $v_c^2 = 2\sigma^2$ (Binney \& Tremaine, 1987). Assuming that the velocity dispersion of the BCG is indicative of the velocity dispersion of the companions in orbit around it near the core of the cluster, we use typical BCG velocity dispersions from literature to estimate $v_c$. Loubser \& Sanchez-Blazquez (2012) find typical BCG velocity dispersion $\sigma\sim 300$~kms$^{-1}$ for nearby BCGs; von der Linden et al. (2007) measure BCGs at $z\sim0.05-0.1$ to have velocity dispersions $\sigma\sim250$~kms$^{-1}$ and similar values are found by Brough et al. (2007). The velocity dispersion of BCGs may evolve with time as the BCG accretes its companions, therefore we calculate $T_{fric}$ using a low and a high estimate for the circular velocity, namely 350 and 700~kms$^{-1}$.

The dynamical timescales for the merging of BCGs and companions at a distance of 50\,kpc with mass ratios $\mathcal{M}_{comp}:\mathcal{M}_{BCG}$ of 1:2, 1:5, 1:10, 1:20 and 1:100 are shown in Table~\ref{tf_tab}. Assuming a $\Lambda$CDM cosmology, the time between $z=0.8$ (the lower redshift limit for the sample used here) and the present is $\sim6.8$\,Gyr ($\sim7.7$\,Gyr for $z=1$). Table~\ref{tf_tab} shows that all the companion galaxies within masses greater than 1/100 of the BCG mass and within 50 kpc of the BCG at $z=1$ should merge with the BCG by $z=0$. 
Any reasonable change in the assumed parameters ($v_c$, M/L) would not significantly alter the merging timescale given by $T_{fric}$. For example, an increase in the circular velocity up to 1000\,kms$^{-1}$ increases the dynamical friction timescale, but only the companions in the lowest mass-ratio bin ($\mathcal{M}_{comp}:\mathcal{M}_{BCG} =$ 1:100) would have merging timescales similar to the cosmic time between $z=0.8$ and the present. 

%Similarly, a decrease in the mass-to-light ratio of the companions to a value as low as  {\bf M/L=2} increases $T_{fric}$ only slightly, and again only the $\mathcal{M}_{comp}:\mathcal{M}_{BCG} = $1:100 companions might be able to escape merging with the BCG by the present. The evolution of the mass-to-light ratio of this type of galaxy between $z=1$ and $z=0$ is {\bf a factor of 0.3} in rest-frame $B$ (van Dokkum \& van der Marel, 2007).
Adopting an evolution in the mass-to-light ratio of a factor of 3 between $z=1$ and $z=0$ (in alignment with van Dokkum \& van der Marel, 2007), a value as low as M/L$=2$ increases $t_{fric}$ such that, again, only the $\mathcal{M}_{comp}:\mathcal{M}_{BCG} = $1:100 companions might be able to escape merging with the BCG by the present.
 The timescales shown in Table~\ref{tf_tab} indicate that all the measured BCG companions at $z\sim1$ in Table~\ref{results_tab} will merge with the BCG by $z=0$.

\begin{table}
\centering
\caption{Dynamical friction timescale for typical masses of companions inferred from their stellar mass ratio the BCG ($\mathcal{M}_{comp}:\mathcal{M}_{BCG}$), using a mass-to-light ratio of M/L=4, and high (700 kms$^{-1}$) and low (350 kms$^{-1}$) estimates for the circular velocity.}
\begin{tabular}{l c c c }
\hline
Stellar mass		& \multicolumn{3}{c}{$T_{fric}$(Gyr)} \\
ratio			&	Low $v_c$	&High $v_c$	& Average\\
\hline
1:2			&0.07	&0.13	&0.10\\
1:5			&0.17	&0.33	&0.25\\
1:10			&0.33	&0.67	&0.50\\
1:20			&0.67	&1.33	&1.00\\
1:100		&3.33	&6.66	&4.99\\
\hline
\end{tabular}
\label{tf_tab}
\end{table}

As previously indicated, to estimate the stellar mass increase of the BCG from the accretion of its companions at $z=1$, we make the implicit assumption that the luminosity ratios of companion to BCG in the observed waveband are directly indicative of their stellar mass ratios (following Edwards \& Patton 2012, whose observations have the same rest-frame waveband as our own). The stellar-mass ratio between a BCG and a companion galaxy is given by their luminosity ($L$) ratio, as described in Equation~\ref{lum_equ}, so that $L_{comp}:L_{BCG} \equiv \mathcal{M}_{comp}:\mathcal{M}_{BCG}$.

\begin{figure}
\centering
\includegraphics[width=8cm]{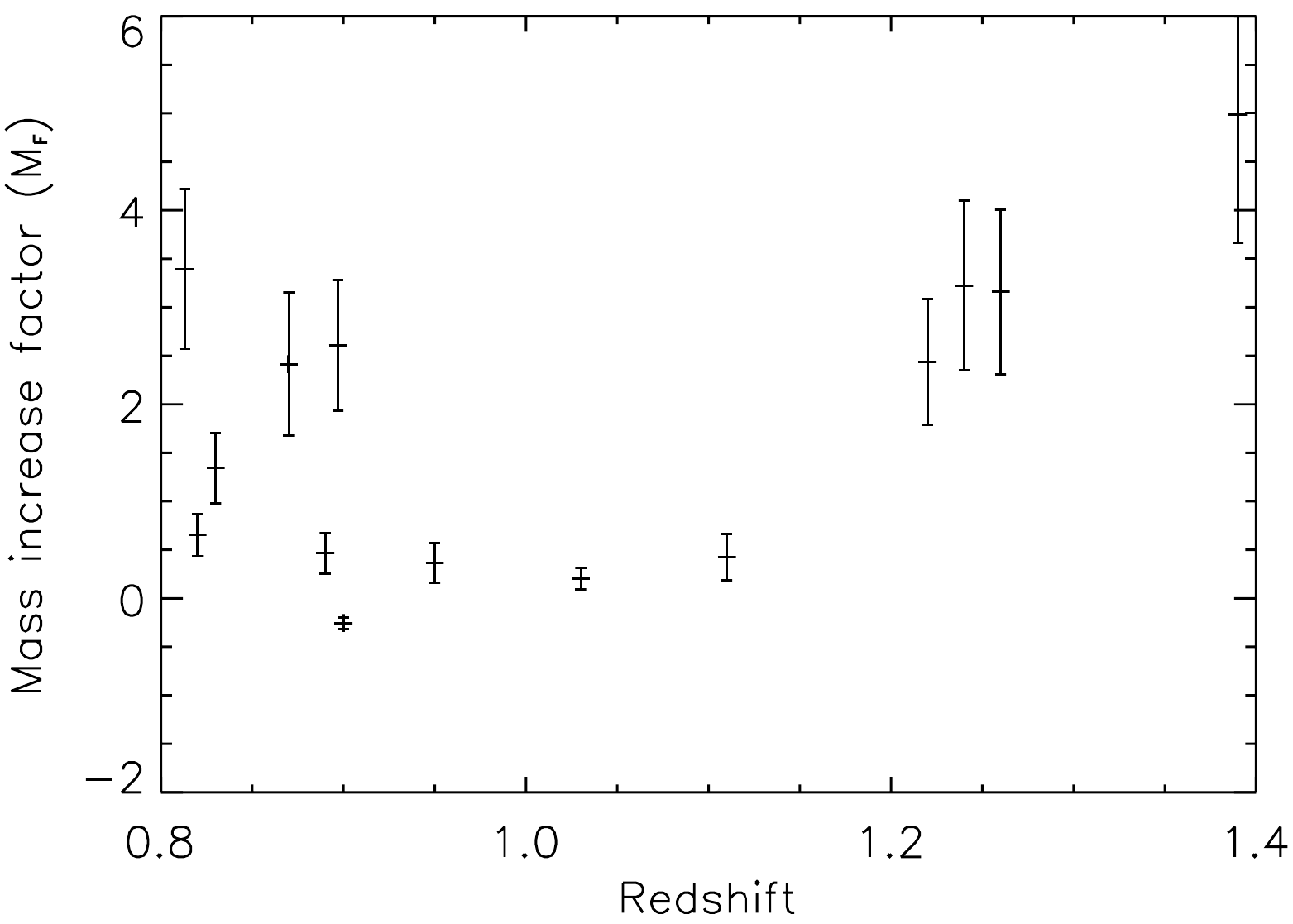}
\caption{Projected stellar mass increase of BCGs vs redshift of the cluster, assuming all companions within a 50 kpc radius at $z=1$ merge by $z=0$.}
\label{m_vs_z}
\end{figure}

To calculate the mass growth factor the number of companions to each BCG in each luminosity bin, $N_c$($L_{comp}:L_{BCG}$) and shown in Table~\ref{results_tab}, is multiplied by the corresponding luminosity fraction for that bin and summed to give the mass increase factor ($\mathcal{M}_f$). To account for possible top-heavy or bottom-heavy bins, we calculate the $\mathcal{M}_f$ for both the minimum and maximum luminosity ratios defining each bin. The mass growth factor ($\mathcal{M}_f$) is then calculated from the average of,
\begin{equation}
\begin{split}
\mathcal{M}_f(min) = 0.5N_c(1:1-1:2) + 0.2N_c(1:2-1:5) \\ + 0.1N_c(1:5-1:10) + 0.05N_c(1:10-1:20),
\end{split}
\label{equ_min}
\end{equation}
and
\begin{equation}
\begin{split}
\mathcal{M}_f(max) = N_c(1:1-1:2) +0.5N_c(1:2-1:5) \\ +0.2N_c(1:5-1:10) + 0.1N_c(1:10-1:20).
\end{split}
\label{equ_max}
\end{equation}
These equations give an average $\mathcal{M}_f=2.15\pm0.43$ using the raw binned counts. Estimating the same quantity by summing over the companion galaxies using their individual mass ratios gives $\mathcal{M}_f=2.00\pm0.31$ (uncorrected), an almost identical value indicating results are not significantly dependent on the choice of binning. Applying the contamination correction using the field average for each bin the predicted stellar mass growth factor for the BCGs since $z\sim1$ is $\mathcal{M}_f=1.83\pm0.43$ (i.e. 183\% larger in mass by $z=0$). The individual stellar mass increase factors are given in Table~\ref{results_tab}, and Figure~\ref{m_vs_z} shows the estimated mass increase for each cluster as a function of redshift. It is worth noting here that these results ignore any contribution from massive companions accreted from outside of a 50\,kpc radius, for which the dynamical friction timescale is less than the available cosmic time between $z=1$ and $z=0$.

\begin{figure}
\begin{center}
\includegraphics[width=8cm]{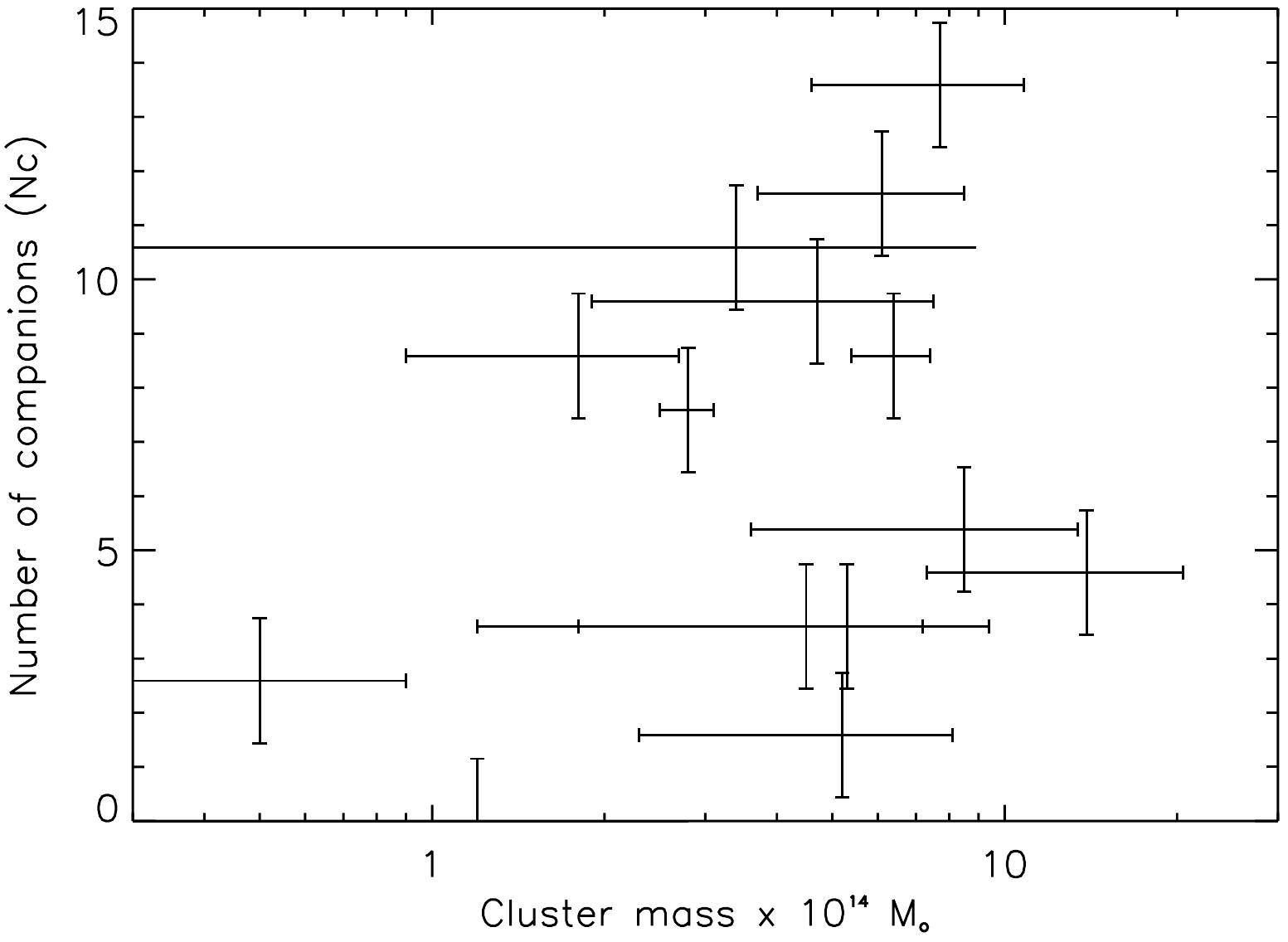}
\caption{Cluster mass vs number of BCG companions for the clusters in our sample.}
\label{nc_vs_cluster_mass}
\end{center}
\end{figure}

\subsection{Caveats}\label{caveats_section}
We emphasise that there are considerable uncertainties associated with our estimates of the mass growth through the accretion of companions onto BCGs at the centres of clusters. For example, our results ignore any contribution from massive companions accreted from outside a 50\,kpc radius, for which the dynamical friction timescale is less than the available cosmic time between $z=1$ and $z=0$, a bias which works in the sense of underestimating our BCG growth rate. This is particularly true for the high mass companions which have a shorter $T_{fric}$. Since the dynamical friction timescale scales as $r^2$ (see Equation~\ref{tf_equ}) and using the results from Table~\ref{tf_tab}, companions with masses 1:5 or larger compared to the BCG, have sufficient time to merge even if located at $\sim$150\,kpc from the cluster centre. Conversely, some of the galaxies measured to be within 50\,kpc of the BCG may not merge due to high velocities as a result of rapid infall into the cluster. 
 
We discuss our results in the context of the build up of the ICL in Section~\ref{implications}, but point out here the consequences of the  process of violent relaxation during the accretion of a companion galaxy, which can cause a substantial fraction of the stars in the accreted companion to be ejected, thereby reducing the available material merging onto the BCG.  The point is clearly made by Murante et al. (2007) who use hydrodynamical simulations to investigate the formation mechanism of the ICL. These suggest that some 50\% of the ICL is associated with the family-merging tree of the BCG, demoting tidal stripping to a more minor role. Furthermore in each significant encounter of a galaxy with the BCG, up to 30\% of the stellar mass becomes unbound and ends up in the ICL, also leading to less mass growth for the BCG. The simulations of Murante et al. also predict that 70\% of ICL forms after $z=1$. When considered in the context our our recent paper (Burke et al. 2012) where we observed large growth of the ICL over the same time, this is suggestive that a large fraction of the mass which would be accreted from the companion galaxies measured in this study will end up in the ICL rather than centrally on the BCG, see Section~\ref{implications}.  

Another bias which may affect our estimated BCG mass growth is the contamination from cluster galaxies which in projection, are within a 50 kpc aperture around the BCG, but are in reality located outside of this radius. In order to examine the magnitude of this effect we carried out the same companion counting on galaxies within 1 magnitude of the BCG in an annulus between 250 and 500 kpc from the cluster centre. For a total of 40 bright galaxies the average number of companions within 50 kpc is $1.64\pm0.18$. Although this is nominally 16\% larger contamination compared with the field-only galaxy counts of $1.41\pm0.19$, reported in Table~\ref{results_tab}, it is still within $1\sigma$. Hence we conclude that we cannot clearly distinguish the difference between background contamination and contamination from cluster galaxies in our data.

\section{Discussion}\label{discussion}

Our results show that for the BCGs at $0.8<z<1.4$ discussed here, the average number of companion galaxies within 50 kpc is $N_c=6.45\pm$1.15. If all of the companions merge with their BCG (as the dynamical friction timescale indicates they should), the BCGs should increase in stellar mass, on average, by a factor of $1.83 \pm0.43$.

The individual $N_c$ values in Table~\ref{results_tab} range from $0-14$ indicating that the `snapshot' images are sampling very different merger rates between the BCGs. In addition the $\mathcal{M}_f$ results for individual clusters shown in Figure~\ref{m_vs_z} indicate a wide range of merging histories for BCGs, with some not undergoing any significant mass increase, while others appear to be destined to grow by a factor  $4-5$. This behaviour is in agreement with the simulations of Laporte et al. (2013). Furthermore whilst the highest $N_c$ values are observed in the highest redshift clusters, the build up of the BCG stellar mass shown in Figure~\ref{m_vs_z} indicates no obvious trend over the redshift range of our sample, and likewise Figure~\ref{nc_vs_cluster_mass} shows no trend between $N_c$ and cluster mass.

\subsection{The role of major and minor mergers}
\begin{figure}
\includegraphics[width=8cm]{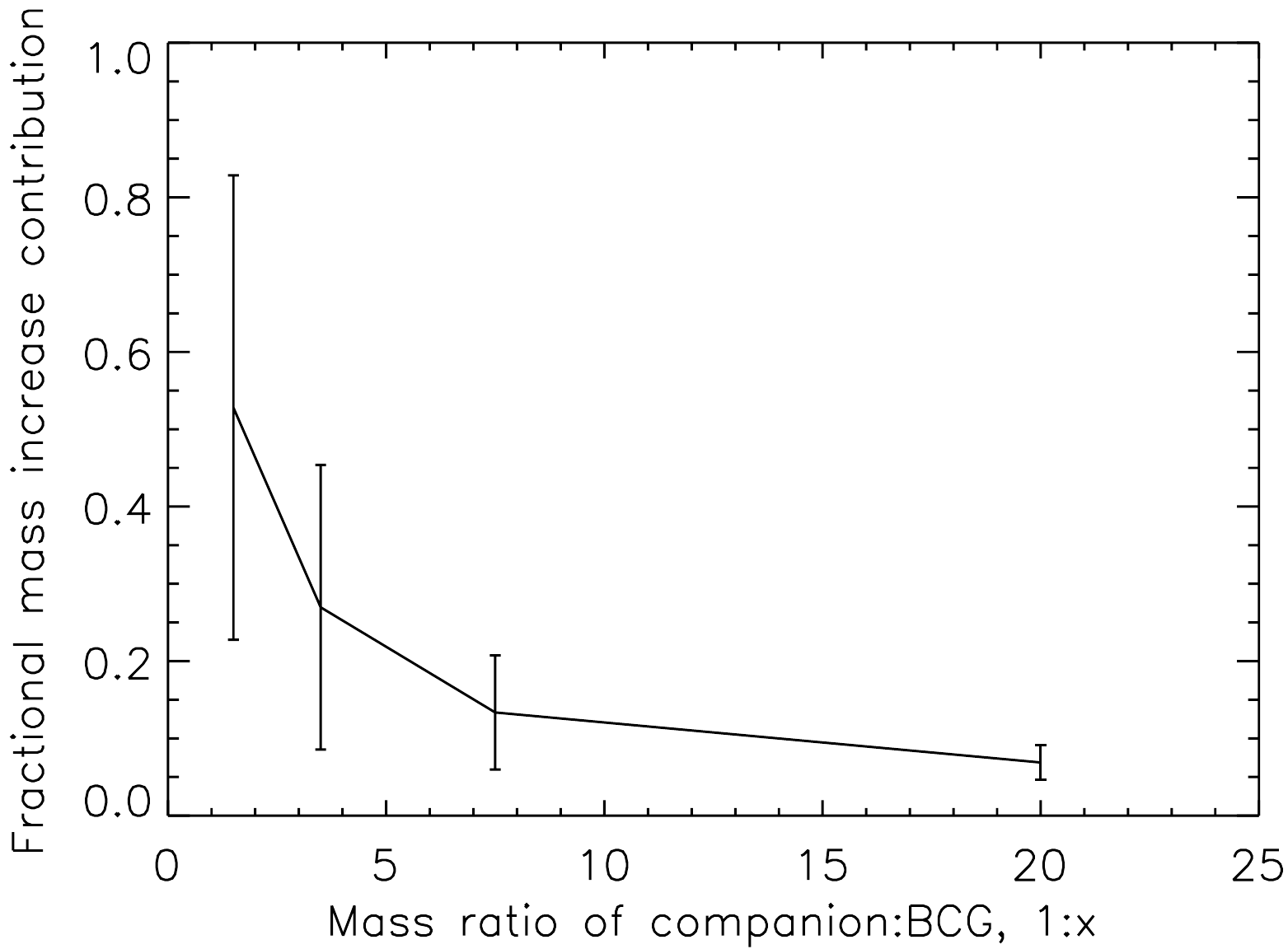}
\caption{The average fractional contribution of companions of given stellar mass ratios ($\mathcal{M}$(comp):$\mathcal{M}$(BCG)) to the average mass growth of the BCG. All values have been normalised to the average total mass growth.}
\label{ratio_contrib}
\end{figure}

 Figure~\ref{ratio_contrib} and Table~\ref{mass_frac_tab} show the contribution to the average mass growth of the BCGs from different mass companions. Approximately 50\% of the mass growth between $z\sim1$ and the present comes from major mergers - companions which have at least half the mass of the BCG (1:1 to 1:2); the rest is made up from minor mergers, with $25\%$ of the BCG mass increase from companions between 1:3 and 1:5 of the BCG mass and a similar contribution from even smaller mass companions. The importance of major mergers is also emphasised in the recent study of bright galaxies in cluster cores by Lidman et al. (2013), who use a complementary method and a sample which contains 5 of the clusters studied here. They also conclude that BCGs in at $z\sim1$ should undergo a large number of major mergers to the present.

\begin{table}
\centering
\caption{Mass increase contributions from major and minor mergers.}
\begin{tabular}{c c }
\hline
Mass ratio  & Fraction of mass \\
$\mathcal{M}_{comp}:\mathcal{M}_{BCG}$ &	growth contributed\\
\hline
1:1~ -- 1:2~	& 0.53 $\pm$ 0.30  \\
1:3~ -- 1:5~	& 0.27 $\pm$ 0.18 \\
1:5~ -- 1:10	& 0.13 $\pm$ 0.07 \\
1:10 -- 1:20	& 0.07 $\pm$ 0.03 \\
\hline

\end{tabular}
\label{mass_frac_tab}
\end{table}

\subsection{Comparison with other observations}\label{comparison}

\begin{figure}
\centering
\includegraphics[width=8cm]{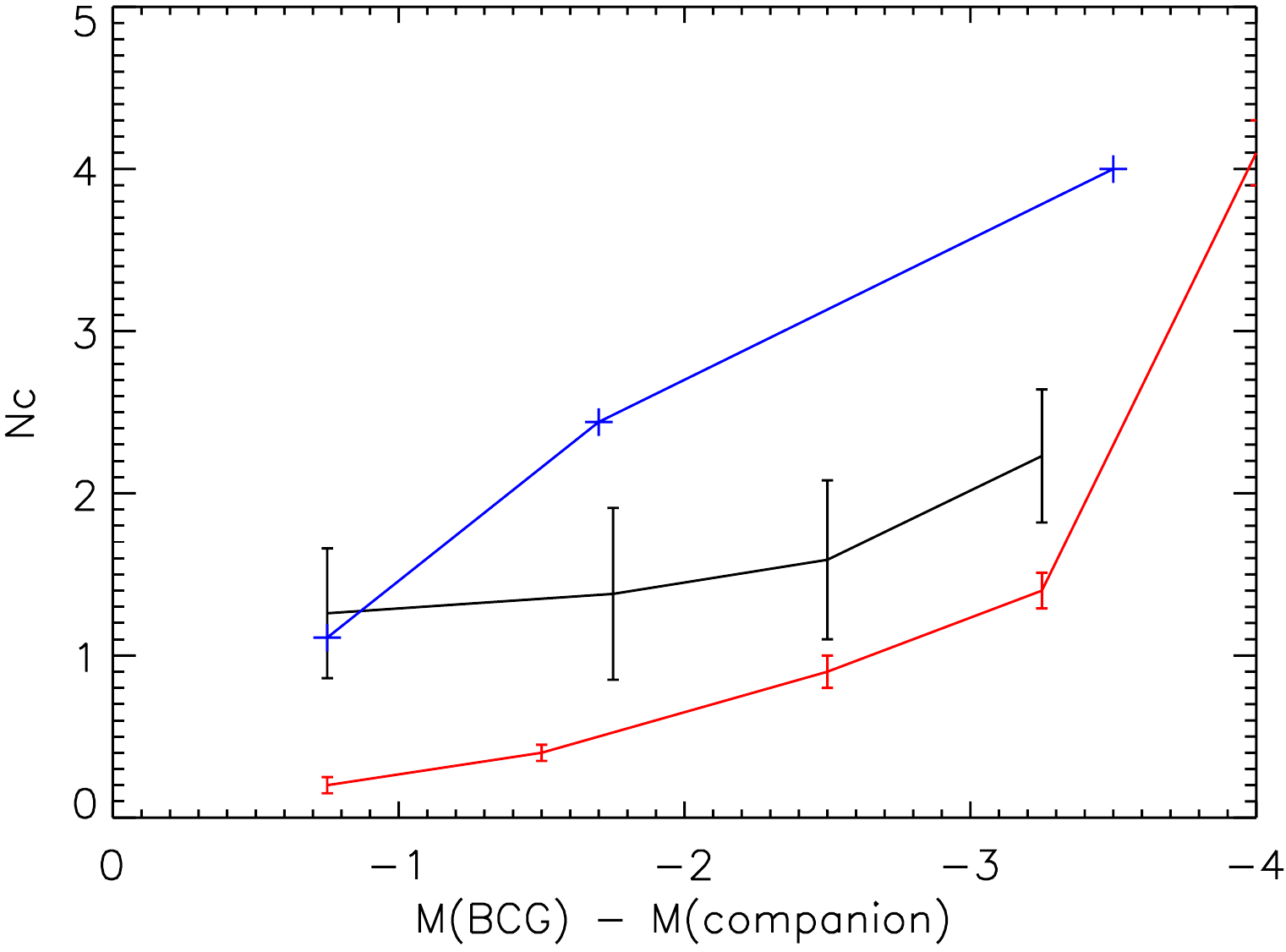}
\caption{Corrected number counts in each magnitude difference (luminosity ratio) bin. The $z\sim1$ sample presented here is shown in black, the results of Edwards \& Patton at $z\sim0.3$ are shown in red and the predictions from the simulations of Laporte et al. (2013) at $z=2$ are shown in blue. The magnitude gaps of -0.75, -1.75, -2.5 and -3.25 correspond to luminosity ratios of 1:2, 1:5, 1:10 and 1:20 respectively.}
\label{ep_comp}
\end{figure}

Edwards \& Patton (2012) study a sample of 91 BCGs at $0.15<z<0.40$ in clusters of masses $>10^{14}M_{\odot}$ with imaging in the $r$-band and carry out a similar analysis to that described here, counting companions within 30 and 50\,kpc apertures and using a control sample of ellipticals of similar brightness to estimate contamination. Using similar dynamical friction arguments they calculate the dynamical merging timescales for the companions in the same manner as described here and conclude that for a 50~kpc aperture $N_c=1.38\pm$0.14 before background correction and $N_c\sim0.7$ after correcting. These are somewhat smaller than our values reported here, namely $N_c=7.86\pm$1.14 and 6.45$\pm$1.15 respectively. A comparison of the $N_c$ values is shown in Figure~\ref{ep_comp}. This figure highlights the excess of companion galaxies to BCGs at $z=1$, with each mass ratio bin containing more than twice the number of companions compared to $z=0.3$. 

In a similar way to the results presented here, Edwards \& Patton find there are a greater number of companions with larger luminosity ratios (20:1) than smaller luminosity ratios (2:1) at $z=0.3$ and in fact find companions of luminosity ratio of 2:1 or smaller  to be rare. This leads them to emphasise the role of  minor merging in the build up BCGs since $z=0.3$ with an estimate that the masses of BCGs increase by $\sim$10\% between $z=0.3$ and $z=0$ from the accretion of their companions. Other observational studies reach similar conclusions: Liu et al (2009) examine 515 BCGs at $z<0.2$ and find $N_c\sim0.1$ for companions with luminosity ratios less than 1:4 and within 30 kpc of the BCG. 
Liu et al. also find 18/515 of their BCGs to be undergoing major mergers and suggest that BCGs should have increased their masses by 15\% from major dry mergers since $z=0.7$. McIntosh et al. (2008) study a sample of BCGs at $z \le 0.12$ and find only 38/845 to be undergoing major mergers ($N_c\sim 0.05$). They predict that massive haloes are growing by $1 - 9\%$  per Gyr by major mergers. Our high redshift sample on the other hand shows larger total luminosities for the 2:1 companions than the 20:1 sample, indicating that major merging is more important for the stellar mass buildup of high redshift BCGs than nearby BCGs. 

In summary, results from the literature find a significantly lower number of BCG companion galaxies in low redshift clusters than are observed here for higher redshift systems. Of course the dynamical timescale argument in Section~\ref{timescale} provides a plausible explanation for this behaviour; Table~\ref{tf_tab}  indicates that all of the BCG companions within 50 kpc and with mass ratios less than 20:1 should have been accreted by the BCG in the 3.4 Gyr between $z=0.8$ and $z=0.3$ and this is consistent with the relatively low numbers of companions found in the $z\sim 0.1 - 0.3$ studies.

\subsection{Comparison with simulations}

Laporte et al. (2013) simulate the evolution of BCGs through their merging history using dark matter haloes initially extracted from the Millennium Simulation (Springel, 2005) and re-simulated using a zoom-in method developed by Gao et al. (2012). They populate the dark matter haloes with stellar particles which are distributed to match the scale-sizes observed for quiescent galaxies at $z=2$ and run their simulation from $z=2 - 0$. The number of companions at each luminosity ratio is compared with our data in Figure~\ref{ep_comp}. On average the predicted instantaneous merger rate over the mass ratios 1:1--2; 1:3--5; 1:10--100  at $z=2$ are a factor 1.5 larger than our observations in the same ratio bins, however our sample has an upper limit of $1.39$ in redshift.

\begin{figure}
\begin{center}
\includegraphics[width=8.5cm]{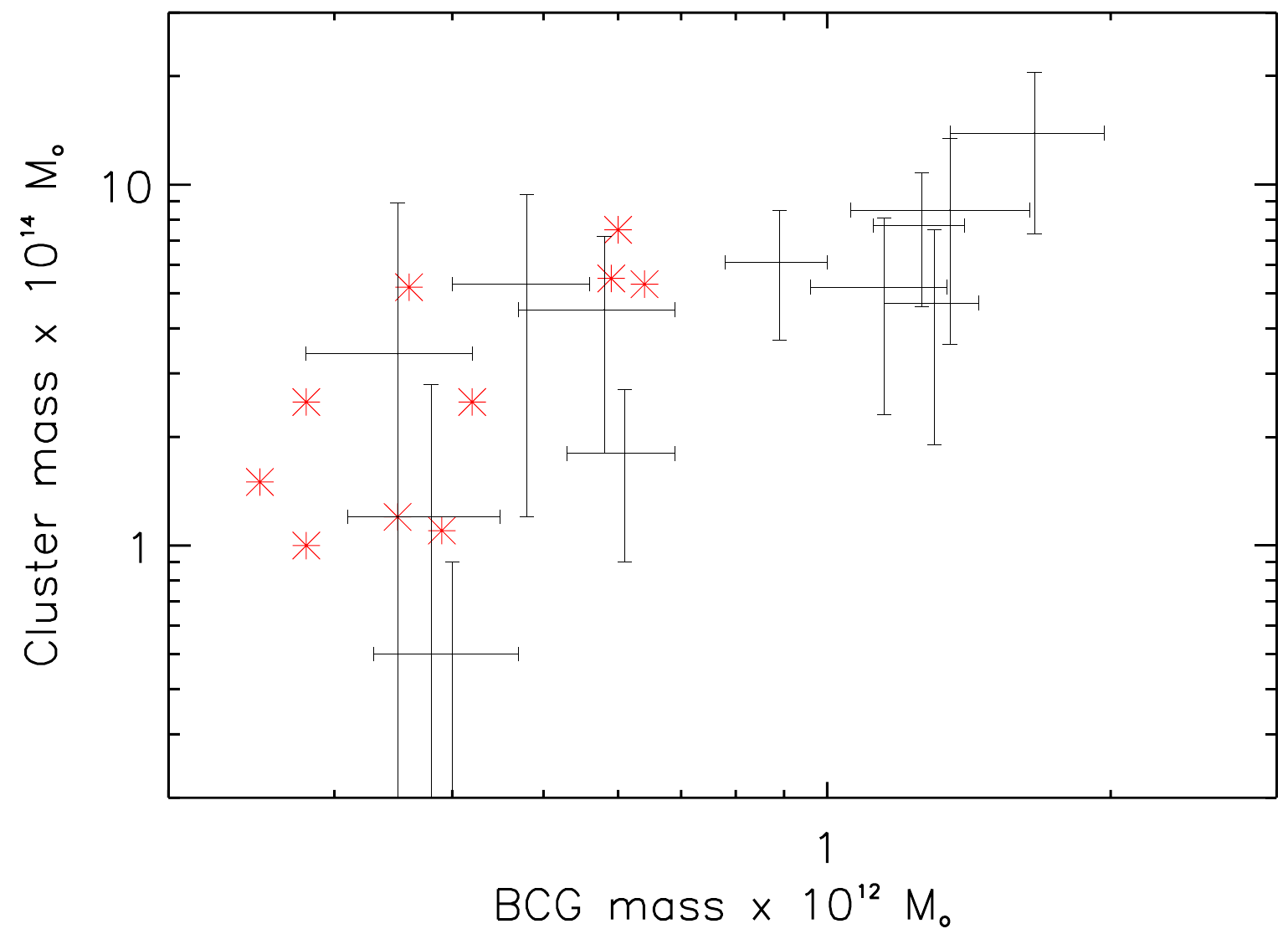}
\caption{The relation between cluster mass and BCG stellar mass for the BCGs in our sample (black) and Laporte et al. (red). The BCG masses for our sample are taken from Stott et al. (2010). This figure excludes RCS~J2319 and RX~J1716 for which we do not have BCG masses.}
\label{mass_comp}
\end{center}
\end{figure}

Although we have no real observational constraint on how the merger rates change with time between $0.3<z<1$ and the uncertainties are large, it is noteworthy that the implied number of mergers during the cosmic time between $z=1-0.3$ is consistent with the major merger rate ($\sim0.4$\,Gyr$^{-1}$) predicted for the most massive galaxies in $\Lambda {\rm CDM}$ simulations (see Hopkins et al., 2010).   

Turning to the BCG stellar mass growth, the Laporte et al. (2013) simulations predict stellar mass growth factors of 1.5 ($0 < z< 0.3$); 1.9 ($0.3 < z < 1$); 2.6 ($0 < z < 1$). Their simulated cluster and BCG masses at $z=1$ are shown in comparison with those of the clusters and BCGs in our sample in Figure~\ref{mass_comp}. From this Figure it can clearly be seen that there is good overlap between the simulations and our observed clusters, reinforcing a valid comparison between the samples. Their predicted mass increase of 1.9 between $0.3 < z < 1$ agrees well with the mass increase measured here under the reasonable assumption (already demonstrated) that all our companions merge by $z=0.3$. However a mass increase factor of 2.6 between $0<z<1$ is outside of the errors on the mass increase found here ($\mathcal{M}_f$ =1.83 $\pm$ 0.43). This appears to be a problem for the mass increase factor of 1.5 predicted for $0<z<0.3$ as the low redshift studies discussed in Section~\ref{comparison} estimate the growth to be at the 10\% level over a similar redshift range. The results of Laporte et al., however, show improved agreement with observations of BCG mass growth at larger redshift over the simulations of De Lucia \& Blaizot (2007), who predict an increase of $3-4$ times in mass for BCGs since $z=1$. However, it is clear from Figure~\ref{mass_comp} that the problem for simulations of large BCG masses at high redshift is still apparent.

\subsection{Implications for stellar mass assembly in cluster cores}\label{implications}

The estimated number of mergers since $z=1$ has direct bearing on the question of whether BCGs have evolved in mass over the same period. As mentioned in the introduction, the mass growth of BCGs measured by some authors since $z=1$ is negligible, whereas others find significant evolution. The growth factor via mergers of $1.8$ reported here is consistent with the stellar mass growth estimate of 1.8 for BCGs between $z=0.9-0.2$ reported by Lidman et al. (2012) but somewhat contradictory to Collins et al (2009) who find 90\% of BCG stellar masses in place by $z \sim1$, and Stott et al (2010) who find BCGs at $z=1$ on average have 95\% of the stellar mass of those measured locally;  both these studies use cluster a sample which largely overlaps with that used here.

If we believe the lower BCG mass growth estimates then it is necessary to ask whether accreted mass from companions ends up in the extended cD halos or as part of the ICL surrounding the BCGs, rather than merging centrally onto the BCG itself. Burke et al. (2012) show that the ICL contains $\sim$4\% of the total cluster light at $z=1$, a fraction which increases to as much as 40\% in clusters nearby (Rudick, Mihos \& McBride 2006, 2011; Gonzalez et al., 2007). Furthermore Burke et al. (2012) estimate that on average the ICL contributes more (observed frame) $J$-band light than the combined stellar emission within a 50~kpc radius of their BCGs  at $z\sim1$ (see their Table 3); providing strong evidence for substantial amounts of light at large radii.  Laporte et al. (2013) do discuss the stripping of galaxies and the amount of mass from mergers which ends up in the diffuse component surrounding the BCG. From their simulations they find between $5-30\%$ of the total merging mass ends up as a ``diffuse'' stripped component. This suggests a scenario where the smallest companion galaxies, which have the longest merging timescales, have their stellar mass slowly stripped as a result of their slow spiral onto the BCG. The drawn-out merger times means that there is ample opportunity for the stars in these small galaxies to be stripped and form the diffuse intracluster light. However, the fraction of the stripped material estimated by Laporte et al. (2013) is smaller than typical estimates of the ICL fraction and there is not enough mass in the companions less than 1/20 of the BCG mass measured in this study (but excluded from our analysis) to account for the large ICL fractions observed in nearby clusters (e.g., Mihos et al., 2005; Rudick et al., 2006; Gonzalez et al., 2005; 2007; Krick et al., 2006; 2007). In addition, the high resolution N-body simulations of Conroy et al. (2007) and Puchwien et al. (2010) predict that $50-80\%$ of the matter accreted onto BCGs from mergers should eventually end up distributed in ICL in order to reproduce the distribution observed in nearby clusters. Furthermore, the simulations of Murante et al. (2007) find that 50\% of the light in the ICL at the present day should have been formed through interactions of cluster galaxies with the BCG. The simulations of the ICL by Rudick et al. (2011) predict a fairly steady buildup of the ICL between $z=2-0$ (see Figure~4 of Rudick et al., 2011), increasing to anything between $9-36\%$ of the total cluster light by today.

Therefore it may reasonably be expected that a smaller fraction of the accreted mass from companion galaxies than is predicted by Laporte et al. and measured in this paper, contributes to the BCG mass growth, with the majority of accreted mass eventually residing in the ICL. In this case, the BCG mass increase from accreted companions reduces from a factor 1.8 (i.e. mass growth of 180\%) to more like $0.5-0.9$ (i.e. mass growth of 50\%--90\%), while the ICL sees a many fold increase in its stellar content between $z=1-0$. More detailed simulations of the BCG + ICL growth at $z=1$ would help to clarify these issues as would more observational work using homogeneously selected cluster samples in the redshift range between $z=0.3-0.8$ and at redshifts higher than $z>1.5$.

Finally, one potential caveat to the BCG evolution work is that the mass estimation from observational measurement almost always assumes that light observed in near-IR wavebands traces stellar mass. In their simulation of BCG mass assembly, Tonini et al. (2012) predict that the use of updated stellar population synthesis models with more detailed treatment of the AGB component can provide reconciliation between the observed no-evolution results and predicted masses of BCGs from simulations. Despite the large scatter in the predictions, this claim should be testable  via the optical to near-IR colours of BCGs at $z>1$.

\section{Conclusions}\label{conc_s}
In this paper we examine the numbers of mergers for 14 BCGs in the redshift range $0.8 < z < 1.4$ using $HST$ images by counting the number of companions to BCGs within 50~kpc apertures to a consistent magnitude limit. Based on these snapshot measurements, BCGs over this redshift range exhibit a wide range of merging activity with an average $\sim6.45\pm1.15$ companions between luminosity or mass ratios 1:1--1:20 after background correction; a number which is significantly larger than similar studies of BCGs at $z=0.3-0.1$. The calculated short merging timescales of the companion galaxies provides an explanation for the smaller number of companions at low redshift and suggests that the majority of the BCG companions at $z=1$ are accreted by $z=0.3$. 

We also estimate the mass growth of the BCGs resulting from the accretion of its companions. We find that major and minor mergers contribute in about equal measure to the mass build of BCGs since $z=1$. The factor by which BCGs are estimated to grow in stellar mass from the accretion of their companions since $z=1$  is $\sim 1.8$, which is close to the predictions of recent simulations of BCG merger history.  If material becomes stripped adding to the build up of the ICL rather than merging onto to BCG over the same period then the real growth of BCGs from mergers could be substantially smaller.

\section*{Acknowledgments}

We thank the anonymous referee for insightful comments which greatly improved this paper and an expeditious report. We thank John Stott for providing helpful discussions.
%IRAF is distributed by the National Optical Astronomy Observatories, which are operated by the Association of Universities Research in Astronomy, Inc., under cooperative agreement with the National Science Foundation.
CAC acknowledges STFC for financial support from grant ST/H/002391/1.
CB acknowledges an STFC quota studentship. 

\newpage
\appendix

\label{lastpage}

\end{document}